\documentclass{JHEP3}
\usepackage[dvips]{graphicx}
\usepackage{amsmath}

\newcommand{\rmd}{\mathrm{d}}
\newcommand{\rmi}{\mathrm{i}}
\newcommand{\rme}{\mathrm{e}}
\newcommand{\D}{\mathcal{D}}
\newcommand{\Lag}{\mathcal{L}}
\newcommand{\Nc}{N_{\text{c}}}
\newcommand{\Nf}{N_{\text{f}}}
\newcommand{\mq}{m_q}
\renewcommand{\Re}{\text{Re}}
\renewcommand{\Im}{\text{Im}}
\newcommand{\1}{{\rm 1\hspace*{-0.4ex}%
\rule{0.1ex}{1.52ex}\hspace*{0.2ex}}}

\title{Characteristics of the eigenvalue distribution
       of the Dirac operator in dense two-color QCD}
\author{Kenji Fukushima\\
        Yukawa Institute for Theoretical Physics\\
        Kyoto University, Kyoto 606-8502, Japan\\
        E-mail: \email{fuku@yukawa.kyoto-u.ac.jp}}
\preprint{YITP-08-45}
\abstract{
 We exposit the eigenvalue distribution of the lattice Dirac operator
 in Quantum Chromodynamics with two colors (i.e.\ two-color QCD).  We
 explicitly calculate all the eigenvalues in the presence of finite
 quark chemical potential $\mu$ for a given gauge configuration on the
 finite-volume lattice.  First, we elaborate the Banks-Casher
 relations in the complex plane extended for the diquark condensate as
 well as the chiral condensate to relate the eigenvalue spectral
 density to the physical observable.  Next, we evaluate the
 condensates and clarify the characteristic spectral change
 corresponding to the phase transition.  Assuming the strong coupling
 limit, we exhibit the numerical results for a random gauge
 configuration in two-color QCD implemented by the staggered fermion
 formalism and confirm that our results agree well with the known
 estimate quantitatively.  We then exploit our method in the case of
 the Wilson fermion formalism with two flavors.  Also we elucidate the
 possibility of the Aoki (parity-flavor broken) phase and conclude
 from the point of view of the spectral density that the artificial
 pion condensation is not induced by the density effect in
 strong-coupling two-color QCD.
}

\begin{document}


\section{Introduction}

  Quantum Chromodynamics with two colors (two-color QCD) instead of
three is a sophisticated practice ground for theorists to extract
worthwhile information out of dense quark matter.  We immediately hit
on several reasons why we can believe so:  First of all, numerous
works on dense two-color QCD have almost established a firm
understanding on the ground state of two-color QCD by the analytical
approach as well as the Monte-Carlo
simulation~\cite{Nakamura:1984uz,Dagotto:1986gw,Dagotto:1986xt,Baillie:1987tr,Klatke:1989xy,Rapp:1997zu,Carter:1998ji,Hands:1999md,Kogut:1999iv,Kogut:2000ek,Hands:2000ei,Splittorff:2000mm,Kogut:2001if,Kogut:2001na,Kogut:2002cm,Splittorff:2002xn,Wirstam:2002be,Muroya:2002ry,Kogut:2003ju,Nishida:2003uj,Ratti:2004ra,Alles:2006ea,Hands:2006ve,Fukushima:2006uv,Fukushima:2007bj,Stephanov:1996ki,Halasz:1997fc,Vanderheyden:2001gx,Akemann:2003wg,Klein:2004hv,Akemann:2006xn}.
Second, the notorious sign problem of the Dirac determinant at
$\mu\neq0$ (where $\mu$ is the quark chemical potential) is not so
harmful as genuine QCD, which makes it viable to perform the
Monte-Carlo
integration~\cite{Nakamura:1984uz,Hands:2000ei,Fukushima:2007bj,Fukushima:2006zz}.
Third, dense two-color matter realizes a bosonic baryon system leading
to the Bose-Einstein condensation of the color-singlet
diquark~\cite{Rapp:1997zu,Carter:1998ji}.  This two-color superfluid
phase is reminiscent of the three-color superconducting
phase~\cite{Alford:2007xm}, for they both break the $\mathrm{U_B}(1)$
symmetry.  Finally, enlarged flavor symmetry earned by the pseudo-real
nature of the SU(2) group that is called Pauli-G\"{u}rsey
symmetry~\cite{Peskin:1980gc,Smilga:1994tb} constrains two-color QCD
at $\mq=\mu=0$.  The interplay between the chiral and diquark sectors
simplifies owing to the symmetry, which enables us to construct an
effective model for two-color QCD with less
ambiguity~\cite{Ratti:2004ra,Fukushima:2007bj}.

  This paper aims to illustrate the spectral behavior in a rather
brute-forth manner.  We usually define the order parameter and concern
its expectation value to examine the phase structure with varying the
external parameters such as the temperature $T$, the quark chemical
potential $\mu$, the quark mass $\mq$, and so on.  We shall explore
our another trail here leading to the phase distinction.  In this work
we will carefully look into the eigenvalue distribution of the Dirac
operator and characterize the state of matter by the distribution
pattern.  Actually, we can find the scatter plot for the Dirac
eigenvalues in two-color QCD in
Refs.~\cite{Baillie:1987tr,Hands:2000ei} and we will do that in a more
systematic way.  It is long known that the eigenvalue spectrum is
informative in the vacuum~\cite{Setoodeh:1988ds} and the random matrix
theory is capable of determining the low-lying spectrum, which has
recently been extended to the finite density
study~\cite{Akemann:2003wg,Akemann:2006xn,Akemann:2007rf}.
Interestingly, the comparison to the random matrix model exhibits good
agreement also in the case of the overlap fermion at
$\mu\neq0$~\cite{Bloch:2006cd}.

  It is not only the low-lying spectrum but also the whole spectral
density that we will deal with in the present paper.  The Monte-Carlo
simulation generates a set of gauge field configurations each of which
has a substantial weight on the partition function.  One configuration
corresponds to one value for a certain operator (the order parameter
for example of our interest), and the more configurations we
accumulate, the more accurately we can improve the expectation value
of the order parameter.  Here, we would remind that the well-known
Banks-Casher relation~\cite{Banks:1979yr} yields the chiral condensate
given in terms of the eigenvalue spectral density at the origin
(i.e.\ $\mq\to0$).  It follows in turn that the order parameter makes
use of only tiny amount of the entire information available from the
spectrum.  In this work, hence, we will unveil detailed information in
a special case of dense and cold ($T=0$) quark matter with two
colors.

  One might come across a question then;  what is the benefit from the
whole shape of the eigenvalue distribution?  To answer this, we should
be aware that the Dirac eigenvalues originally lie on the imaginary
axis except for the displacement in the real direction by $\mq$ but
they scatter over the complex plane because of non-zero chemical
potential $\mu\neq0$ or Wilson coefficient $r\neq0$.  This feature
has, more or less, something to do with the sign problem meaning that
the Dirac determinant could take a negative value.  For $\mu\neq0$ the
Dirac operator mixes the Hermitean and anti-Hermitean operators up
resulting in a complex eigenvalue.  The situation at finite $\mu$
looks similar to that in the presence of the Wilson term in view of
the eigenvalues particularly in the two-color
case~\cite{Fukushima:2006zz}.  [We implicitly assume only the
two-color case below.]  The sign problem may arise actually when the
eigenvalue distribution protrudes from the positive quadrant into
the negative quadrant.  This observation implies that a large $\mq$
(center location of the eigenvalue distribution) as compared to the
chemical potential or the Wilson term (distribution width) would put
the sign problem away.  In the physics language, the vacuum stays
empty as long as $\mq>\mu$, and so there is no $\mu$ dependence
then, which brings about no sign problem naturally.  The onset of the
density effect is manifestly visible from the whole eigenvalue
distribution.  Besides, since $\mq\neq0$ shifts the distribution, it
is transparent to take account of the mass effect provided that the
eigenvalue distribution is given.  These motivate us to turn to the
entire Dirac spectrum.  In the future, hopefully, we believe that the
eigenvalue distribution should shed light upon the sign problem at a
deeper level.  In fact, as we will recognize later, a large value of
$\mu$ induces a peculiar structure in the eigenvalue pattern.

  The above mentioned may well be somewhat abstract.  Let us then make
the issue to be discussed more specific.  What puzzles us is that
there seems to be no clear distinction between the onset criterion
for the superfluid phase and the Aoki phase if considered based on the
eigenvalue distribution alone.  They can possibly coexist but it would
be a weird situation because the superfluid phase is a physical ground
state but the Aoki phase is a lattice artifact inherent in the Wilson
fermion formalism~\cite{Aoki:1983qi,Aoki:1987us}.  The final part of
this paper will be devoted to resolving this matter.  There, we will
find that the onset criterion is certainly degenerate when $\mu=0$,
but the Aoki phase is taken over by the superfluid state in the proper
limit of $\mu\to0$.  In short, we conclude that the Aoki phase never
emerges by the density effect in strong-coupling two-color
QCD.\ \ This statement does not conflict the preceding strong coupling
analysis~\cite{Aoki:1983qi,Aoki:1987us} because the Aoki phase
solution at strong coupling is a saddle-point and infinite $\Nc$ is
required for stability, though this fact is sometimes overlooked.


\section{Two-Color QCD at Strong Coupling}

  In the limit of the strong coupling the gauge action does not enter
the dynamics and the partition function is simply given by the
fermionic part;
\begin{equation}
 Z = \Bigl\langle (\det\D)^{\Nf} \Bigr\rangle_U
  \equiv \int\prod_{n,\mu}\rmd U_\mu(n)\,
  \bigl(\det\D\bigr)^{\Nf} .
\end{equation}
Here $\D$ is the Dirac operator.  Although the strong coupling limit
is a drastic approximation which neglects the gauge dynamics
completely, it is amazing that only the Dirac determinant with random
gluon fields can grasp rich contents of quark matter not only in the
two-color case ~\cite{Dagotto:1986gw,Klatke:1989xy,Nishida:2003uj} but
also in the general
case~\cite{Dagotto:1986xt,Aoki:1983qi,Aoki:1987us,Kawamoto:1981hw,Hoek:1981uv,KlubergStern:1982bs,KlubergStern:1983dg,Fukushima:2002ew,Fukushima:2003fm,Fukushima:2003vi,Kawamoto:2005mq}.

  In a box with volume $V=L^4$, the operator $\D$ is a
$(4\Nc V)\times(4\Nc V)$ matrix.  We denote the eigenvalue of $\D$ by
$\lambda_i$, that is,
\begin{equation}
 \D v_i = \lambda_i v_i \,,
\end{equation}
where $i$ runs from $1$ to $4\Nc V$.  Then the Dirac determinant is
given by the product of all the eigenvalues.  It is easy to prove that
$\det\D$ in the SU(2) gauge theory takes a real value even at finite
density where $\D$ loses the $\gamma_5$-Hermiticity,
i.e.\ $\gamma_5\D(\mu)\gamma_5=\D^\dagger(-\mu)\neq\D^\dagger(\mu)$.
The standard argument immediately follows;
\begin{equation}
 \det\D(\mu) = \det\bigl[(C\sigma_2\gamma_5)^{-1}\D(\mu)
  (C\sigma_2\gamma_5)\bigr] = \det\D^\ast(\mu)
  = \bigl[ \det\D(\mu) \bigr]^\ast \,.
\end{equation}
Here, to derive the above, the necessary relations we use are
$\gamma_5\gamma_\mu\gamma_5=-\gamma_\mu$,
$C\gamma_\mu C^{-1}=-\gamma^T_\mu$, and $\sigma_2 U\sigma_2=U^\ast$
where the last relation corresponds to the pseudo-real nature of the
SU(2) group.

  From this argument we see that $\det\D(\mu)$ is real but not
necessarily positive.  The simulation thus entails an even number
of $\Nf$ so that $(\det\D)^{\Nf}$ is positive definite.  This is the
main reason why the exotic phase structure proposed in
Refs.~\cite{Splittorff:2000mm,Fukushima:2007bj} in two-color QCD with
quark and isospin chemical potentials has been far from confirmed.
The two-color determinant, however, buries a nice property of
respective eigenvalues under the product.  We can prove that, if
$\lambda_i=\mq+\rmi\lambda_i'$ is an eigenvalue of the Dirac
determinant in two-color QCD, there appear $\mq-\rmi\lambda_i'$,
$\mq+\rmi\lambda_i^{\prime\ast}$, and $\mq-\rmi\lambda_i^{\prime\ast}$
simultaneously in the eigenvalue
spectrum~\cite{Hands:2000ei,Fukushima:2007bj,Fukushima:2006zz}.  The
proof may break down when $\rmi\lambda_i'$ is a real number;  the
eigenvectors for $\mq+\rmi\lambda_i'$ and
$\mq-\rmi\lambda_i^{\prime\ast}$ could not be independent.  According
to Ref.~\cite{Hands:2000ei} the staggered fermion is safe from such a
possibility but the Wilson fermion has only a pair of
$\mq+\rmi\lambda_i'$ and $\mq-\rmi\lambda_i'$ instead of a complex
quartet in that case of real $\rmi\lambda_i'$.  We will explicitly
verify that this is the case.  Then the single-flavor Wilson fermion
suffers the sign problem once either of real $\mq+\rmi\lambda_i'$ and
$\mq-\rmi\lambda_i'$ is negative.

  Here we shall briefly summarize the known facts in two-color QCD at
strong coupling.  Let us begin with the chiral limit.  It has been
discovered first in Ref.~\cite{Dagotto:1986gw} that the chiral
condensate is zero, while the diquark condensate has a finite
expectation value, in the limit of $\mu\to0$ with $\mq=0$ taken
first.  In the presence of $\mq\neq0$ the system is kept intact as
long as $\mu$ is sufficiently small and, in turn, the chiral
condensate becomes non-zero but the diquark condensate vanishes.  As
soon as $\mu$ exceeds the mass of the lightest excitation (usually
bosonic baryon), the density effect is activated leading to decreasing
chiral condensate and increasing diquark condensate as $\mu$ goes
larger.  We remark that this behavior of dense two-color QCD has been
settled in the staggered fermion but the relation between the chiral
and diquark condensates is not quite convincing yet in the Wilson
fermion because the Wilson term breaks chiral symmetry explicitly.


\section{Banks-Casher Relations}

  Here we will make a quick view over the link between the eigenvalue
spectral density and the chiral, diquark, and parity-flavor breaking
condensates for later usage.  In this section the argument holds
regardless of strong coupling or not.


\subsection{Chiral Condensate}

  It is widely known that the chiral condensate has a close connection
to the Dirac eigenvalue distribution via the Banks-Casher
relation~\cite{Banks:1979yr}.  To advance our discussions in a
self-contained manner we shall take a brief look at the derivation of
the Banks-Casher relation.  In the explicit presence of the source for
the chiral condensate (i.e.\ mass term), the Dirac operator could be
decomposed into the form of $\D[m]=\mq\1+\D[0]$ whose eigenvalue is
denoted as $\lambda_i=\mq+\rmi\lambda_i'$ as we did in the previous
section.  The chiral condensate per flavor is given by the derivative
of $Z$ with respect to $\mq$, which leads us to
\begin{equation}
 \begin{split}
 \frac{1}{\Nf}\langle\bar{\psi}\psi\rangle &= -\frac{1}{\Nf V}
  \frac{\partial}{\partial m}\ln Z
  = -\frac{1}{V}\Bigl\langle \sum_i \frac{1}{\lambda_i} \prod_j
  \lambda_j \Bigr\rangle_U\cdot
  \Bigl\langle\prod_j\lambda_j\Bigr\rangle_U^{-1} \equiv\\
 &\equiv -\frac{1}{V}\biggl\langle\!\!\biggl\langle \sum_i
  \frac{1}{\lambda_i} \biggr\rangle\!\!\biggr\rangle
  = \biggl\langle\!\!\biggl\langle
  \oint\frac{\rmd\lambda}{2\pi\rmi}\,\frac{\pi\rho_\chi(\lambda)}
  {\lambda} \biggr\rangle\!\!\biggr\rangle \,,
 \end{split}
\end{equation}
where $\rho_\chi(\lambda)$ is the eigenvalue spectral density which is
to be expressed in the complex plane as
\begin{equation}
 \rho_\chi(\lambda) \equiv \frac{1}{\pi V}\sum_i
  \frac{1}{\lambda_i-\lambda} \,,
\label{eq:density}
\end{equation}
which is, strictly speaking, the resolvent~\cite{Akemann:2007rf}
rather than the spectral density.  To keep the analogy to the
conventional Banks-Casher relation, however, we shall refer to the
above as the spectral density.  The integration contour should go
around all of the poles at $\lambda_i$ to pick all the eigenvalues up.
In our notation $\langle\cdots\rangle_U$ means the ensemble average
over gauge configurations and $\langle\!\langle\cdots\rangle\!\rangle$
represents the average including the Dirac determinant.

  Here we consider the contour which is an infinitely large circle in
the complex plane surrounding all the poles.  Then the contour
integral must amount to zero because $\rho_\chi(\lambda)/\lambda$ goes
to zero faster than $|\lambda|^{-1}$.  That means that we can evaluate
the above integral by the negative residue of the pole at
$\lambda=0$.  After all, we have
\begin{equation}
 \langle\bar{\psi}\psi\rangle = -\Nf\,\pi\bigl\langle\!\bigl\langle
  \rho_\chi(0)\bigr\rangle\!\bigr\rangle \,.
\label{eq:Banks-Casher}
\end{equation}
For consistency check let us consider a bit more about this formula.
Usually the Banks-Casher relation is given in the limit of $\mq\to0$.
In this limit, using the notation $\lambda_i=\mq+\rmi\lambda_i'$ where
$\lambda_i'$ is a real number in the continuum theory, we can rewrite
Eq.~(\ref{eq:density}) into a form of
\begin{equation}
 \rho_\chi(0) = \lim_{\mq\to0}\frac{1}{\pi V}\sum_i
  \frac{1}{\mq+\rmi\lambda_i'}
  = \frac{1}{V}\sum_i \delta(\lambda_i') \,,
\label{eq:delta}
\end{equation}
which is more familiar in literatures.  We note that
Eq.~(\ref{eq:density}) is in fact an analytic continued form of the
expression~(\ref{eq:delta}) with the delta function, and it is
equivalent to the definition of the resolvent used in the context of
the random matrix theory~\cite{Akemann:2003wg}.  This complex
extension is necessary for our purpose since the Dirac operator loses
Hermiticity at finite density or in the Wilson fermion formalism.  One
might have noticed that the Banks-Casher relation in the complex plane
is a trivial relation; it is obvious from Eq.~(\ref{eq:density}) that
$-\pi\rho_\chi(0)$ returns to $\sum_i\lambda_i^{-1}$ immediately.


\subsection{Diquark Condensate}

  We can develop the same argument for the diquark condensate as well
as the chiral condensate.  We shall limit our discussions to the case
with degenerate two-flavor ($u$ and $d$) quarks, and then we do not
need to introduce the Nambu-Gor'kov basis.  In the presence of the
same quark chemical potential $\mu$ for $u$ and $d$ quarks, we can
write the Lagrangian density down as~\cite{Hands:2006ve}
\begin{equation}
 \Lag = \bar{\psi}_u\D(\mu)\psi_u + \bar{\psi}_d\D(\mu)\psi_d
  - J\bar{\psi}_u(C\gamma_5)\sigma_2\bar{\psi}_d^T
  + \bar{J}\psi_d^T(C\gamma_5)\sigma_2\psi_u \,,
\end{equation}
where $J$ and $\bar{J}$ are the source for the diquark and
anti-diquark which are anti-symmetric in spin, color, and flavor.  By
means of a variable change by
\begin{equation}
 \bar{\phi}_d \equiv \psi_d^T C\sigma_2 \,,\quad
 \phi_d \equiv C\sigma_2 \bar{\psi}_d^T \,,
\end{equation}
it is possible to compactify the above into
\begin{equation}
 \Lag = (\bar{\psi}_u, \bar{\phi}_d) \begin{pmatrix}
  \D(\mu)\; & -J\gamma_5 \\ \bar{J}\gamma_5\;\; & \D(-\mu)
  \end{pmatrix}
 \begin{pmatrix} \psi_u \\ \phi_d \end{pmatrix} \,.
\end{equation}
The integration over the quark fields is then straightforward and the
resultant partition function is given as the determinant as follows;
\begin{equation}
 Z(J) = \Biggl\langle\det\begin{pmatrix}
  \D(\mu)\gamma_5 & -J \\ \bar{J} & \D(-\mu)\gamma_5 \end{pmatrix}
  \Biggr\rangle_U = \Bigl\langle \det\Bigl[ \D(\mu)\D^\dagger(\mu)
  + |J|^2 \Bigr] \Bigr\rangle_U \,,
\label{eq:par_J}
\end{equation}
where we have used $\gamma_5\D(-\mu)\gamma_5=\D^\dagger(\mu)$.  We
note that $\D(\mu)\D^\dagger(\mu)$ is always Hermitean though
$\D(\mu)$ may not be so.  We can then prove that the eigenvalue of
$\D(\mu)\D^\dagger(\mu)$ is non-negative real, which we denote by
$\xi_i^2$ with choosing $\xi_i\ge0$.  The diquark condensate thus
reads
\begin{equation}
 \bigl\langle\bar{\psi}_u(C\gamma_5)\sigma_2\bar{\psi}^T_d
  \bigr\rangle = \frac{\partial}{V\partial J}Z(J)\Bigr|_{J=0}
  = \frac{1}{V}\biggl\langle\!\!\biggl\langle \sum_i
  \frac{J}{\xi_i^2 + |J|^2} \biggr\rangle\!\!\biggr\rangle
  = \pi\bigl\langle\!\bigl\langle
  \rho_D(0)\bigr\rangle\!\bigr\rangle \,,
\label{eq:Banks-Casher_d}
\end{equation}
where we have defined the diquark spectral density,
\begin{equation}
 \rho_D(\xi) = \frac{1}{V}\sum_i \delta(\xi-\xi_i) \,,
\end{equation}
in a familiar form.  It should be mentioned that we do not have to
perform the analytic continuation this time because $\xi_i$ sits on
the real axis.


\subsection{Parity-Flavor Breaking Condensate}

  In the same way we can discuss the parity-flavor breaking condensate
whose non-zero expectation value characterizes the Aoki phase in the
Wilson fermion formalism.  For two-flavor quarks the source term for
the condensate $\langle\bar{\psi}\rmi\gamma_5\tau_3\psi\rangle$ enters
the Lagrangian as
\begin{equation}
 \Lag = \bar{\psi}_u\D(\mu)\psi_u +\bar{\psi}_d\D(\mu)\psi_d
  + H(\bar{\psi}_u\rmi\gamma_5\psi_u - \bar{\psi}_d\rmi\gamma_5\psi_d)
  \,,
\end{equation}
from which the partition function reads
\begin{equation}
 Z(H) = \Biggl\langle \det\begin{pmatrix} \D(\mu)\gamma_5
  + \rmi H & 0 \\ 0 & \D(\mu)\gamma_5 - \rmi H \end{pmatrix}
  \Biggr\rangle_U = \Bigl\langle \det\Bigl[ \D(\mu)\D^\dagger(-\mu)
  + H^2 \Bigr] \Bigr\rangle_U \,.
\label{eq:par_H}
\end{equation}
It is interesting to note that Eq.~(\ref{eq:par_H}) above is reduced
to Eq.~(\ref{eq:par_J}) when $\mu=0$.  As a result the parity-flavor
breaking condensate seems to be degenerate with the diquark
condensate in the absence of chemical potential.  Once the finite
density is switched on, $D(\mu)D^\dagger(-\mu)$ is no longer a
Hermitean operator, and its eigenvalue distribution spreads over the
complex plane.  Thus, if we define the spectral density by
\begin{equation}
 \rho_H(\eta) \equiv \frac{1}{\pi V}\sum_i
  \frac{1}{\eta_i - \eta}
\end{equation}
with the complex eigenvalue $\eta_i^2$ of the operator
$\D(\mu)\D^\dagger(-\mu)$ with choosing $\Re(\eta_i)\ge0$, we can
write the parity-flavor breaking condensate as
\begin{equation}
 \bigl\langle\bar{\psi}_u\rmi\gamma_5\psi_u
  -\bar{\psi}_d\rmi\gamma_5\psi_d\bigr\rangle = -\rmi\pi\bigl\langle
  \!\bigl\langle \rho_H(\rmi H)-\rho_H(-\rmi H) \bigr\rangle\!
  \bigr\rangle = 2\pi\Im\bigl\langle\!\bigl\langle
  \rho_H(\rmi H) \bigr\rangle\!\bigr\rangle \,.
\end{equation}


\section{Eigenvalue Distribution for a Random Configuration}

  In this work we will take only one random configuration as a
representative instead of calculating the ensemble average over many
random configurations.  Actually the eigenvalue distribution for one
typical gauge configuration turns out to be quite informative in our
case.  This simplification is legitimate because each random
configuration equally contributes to a physical quantity in the strong
coupling limit.  So to speak, the strong coupling theory is democratic
and any configuration is eligible for a representative.  If we are
interested in the weak coupling regime, we would have to take an
appropriate ensemble average.

  We will first proceed to the calculation in the staggered fermion
formalism and make sure that our results agree well with known results
in the mean-field approximation at strong coupling.  After that we
will adopt the Wilson fermion formalism and look further into the
possibility of the Aoki phase.


\subsection{Staggered Fermion}

  The Dirac operator at finite density in the staggered fermion
formalism is
\begin{equation}
 \begin{split}
 \D_S(\mu) &\equiv \mq\,\delta_{m,n} + \frac{1}{2}\sum_i \eta_i(m)\Bigl[
  U_i(m)\,\delta_{m+\hat{\imath},n} -U^\dagger_i(n)\,\delta_{m,n+\hat{\imath}}
  \Bigr] +\\
 &\quad + \eta_4(m)\Bigl[ \rme^\mu\,U_4(m)\,\delta_{m+\hat{4},n}
  -\rme^{-\mu}\,U^\dagger_4(n)\,\delta_{m,n+\hat{4}} \Bigr] \,,
 \end{split}
\end{equation}
where $\eta_\mu(n)\equiv(-1)^{n_1+n_2+\cdots+n_{\mu-1}}$ and the
chemical potential is introduced as formulated in
Ref.~\cite{Hasenfratz:1983ba}.

\FIGURE{
\includegraphics[width=12cm]{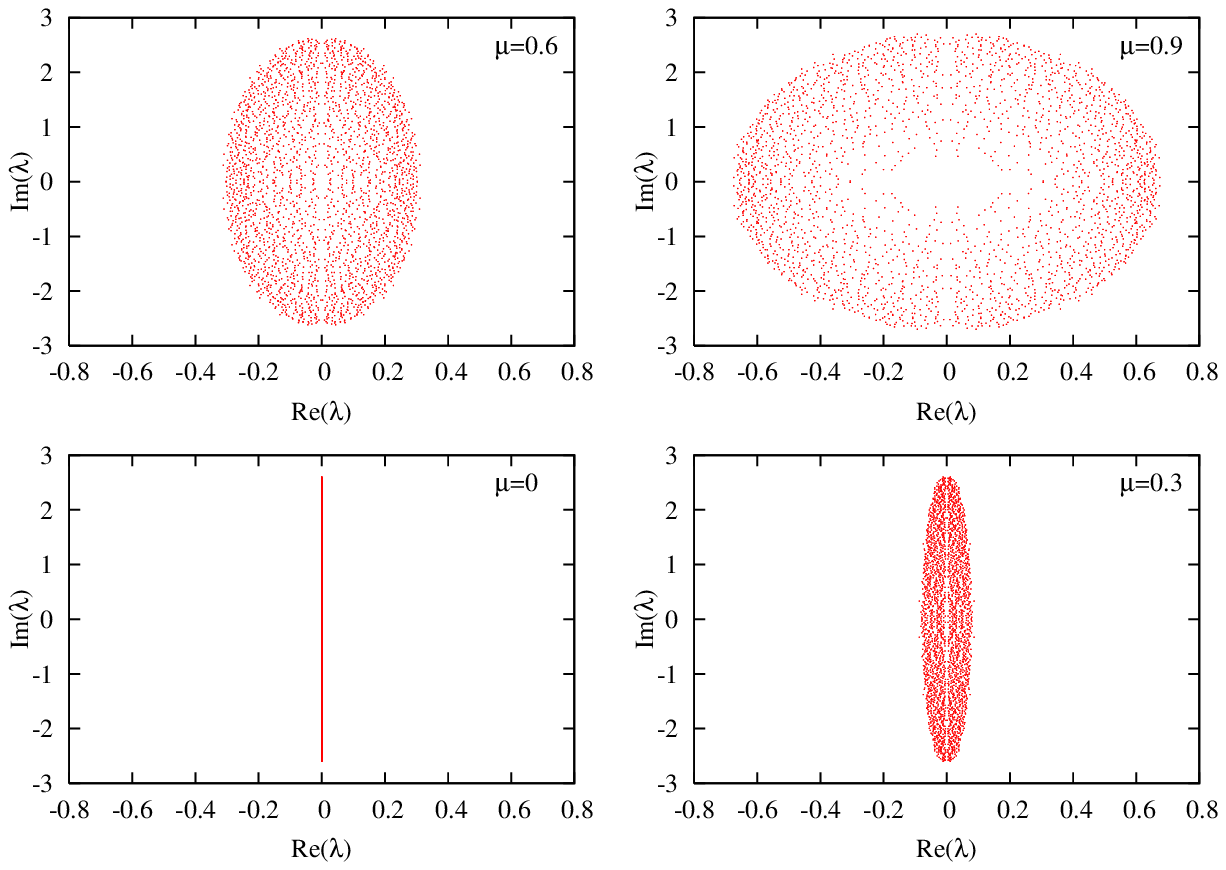}
\caption{Eigenvalue distribution for a random gauge configuration in
  the staggered fermion formalism at $\mq=0$ on the $6^4$ lattice.
  The distribution at $\mq\neq0$ is given by a shift along the real
  axis by $\mq$.}
\label{fig:staggered}
}

  The zero-density Dirac operator in the staggered fermion formalism is
anti-Hermitean except for the mass term, so that all the eigenvalues
reside on the vertical line whose real part is $\mq$ (see the
lower-left scatter plot in Fig.~\ref{fig:staggered}).  The chemical
potential breaks Hermiticity and the eigenvalue distribution has a
width along the real axis as $\mu$ goes larger as shown in the
scatter plots for $\mu=0.3$, $0.6$, and $0.9$ in
Fig.~\ref{fig:staggered}.  These figures are reminiscent of the
scatter plots in Refs.~\cite{Baillie:1987tr,Hands:2000ei}.  To draw
Fig.~\ref{fig:staggered} we have generated a random gauge
configuration on the lattice with a volume of $V=6^4$.  Because the
staggered fermion has only color indices, the total number of dots in
Fig.~\ref{fig:staggered} is $2\times6^4=2592$ for each plot.  We have
made use of LAPACK to compute $2592$ eigenvalues numerically.

  The broadened width in the real direction has a definite physical
meaning.  In the case of $\mq\neq0$ the distribution has to be shifted
by $\mq$ and then the entire eigenvalue distribution can be placed in
the positive quadrant as long as $\mu$ is small as compared to $\mq$.
It is hence a natural anticipation that the superfluidity has an onset
when the eigenvalue distribution becomes as wide as it reaches the
origin.  This is actually the case.

\FIGURE{
\includegraphics[width=12cm]{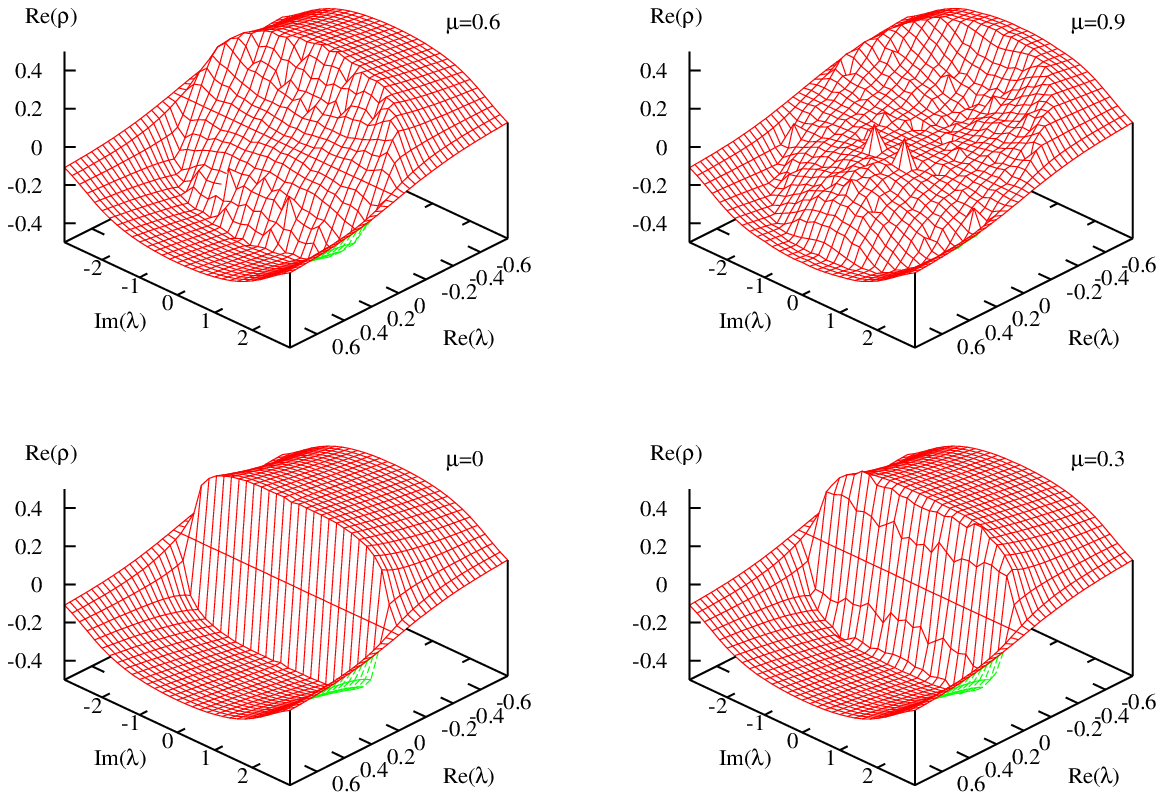}
\caption{Real part of the spectral density, $\Re(\rho_\chi(\lambda))$,
  in the complex plane for various values of the chemical potential.}
\label{fig:chiral_st}
}

  From the obtained eigenvalues we can explicitly calculate the
spectral density~(\ref{eq:density}) to evaluate the chiral condensate
through the Banks-Casher relation in Eq.~(\ref{eq:Banks-Casher}).
Because of the quartet pattern of the eigenvalue distribution the
imaginary part of $\rho_\chi(\lambda)$ is vanishing on the real axis.
The chiral condensate inferred from Eq.~(\ref{eq:Banks-Casher}) is
thus insensitive to the imaginary part but determined solely by the
real part of the spectral density taking a real value.  We show the
real part of the spectral density~(\ref{eq:density}) in
Fig.~\ref{fig:chiral_st} for various $\mu$.  It is remarkable that the
spectral density for a random configuration looks such smooth even
without taking an ensemble average.

  As we have mentioned, the eigenvalues and thus the spectral density
with a finite $\mq$ can be deduced simply by a shift along the real
axis by $\mq$.  Therefore, $\rho_\chi(0)$ appearing in
Eq.~(\ref{eq:Banks-Casher}) can be read from Fig.~\ref{fig:chiral_st}
by the value at $(\Re\lambda,\Im\lambda)=(-\mq,0)$.

  When $\mu=0$ a sharp perpendicular wall stands at $\Re(\lambda)=0$
which is responsible for a non-vanishing chiral condensate in the
limit of $\mq\to0$ while keeping $\mu=0$.  The wall is smoothened by
the effect of $\mu\neq0$ and it is no longer vertically upright at
finite density, which leads to an interesting observation.  In fact,
it is not hard to conceive from Fig.~\ref{fig:chiral_st} that the
chiral condensate becomes zero in the chiral limit at infinitesimal
but nonzero $\mu$.  This is absolutely consistent with
Ref.~\cite{Dagotto:1986gw}.

  We shall next evaluate the diquark condensate using the Banks-Casher
relation~(\ref{eq:Banks-Casher_d}).  We will start with the chiral
limit ($\mq=0$) and then go into the finite mass case that we choose
$\mq=0.2$ here in this work.  For convenience we define the integrated
diquark spectral number,
\begin{equation}
 n_D(\xi) = \int_0^\xi \!\rmd\xi'\, \rho_D(\xi') \,,
\end{equation}
whose slope at $\xi=0$ gives the spectral density $\rho_D(\xi=0)$
which is proportional to the diquark condensate.  Although the
staggered fermion Lagrangian does not involve the Dirac spinor, it is
not difficult to make use of the Nambu-Gor'kov representation to
express the diquark condensate by the diquark spectral density.  Since
the derivation is only straightforward, we will not reiterate it but
skip detailed arithmetics.  To summarize the resultant relations, we
can prove that
\begin{equation}
 \sigma \equiv \frac{1}{2}\langle\bar{\chi}\chi\rangle
  = \frac{\pi}{2}\rho(0) \,,\qquad
 \Delta \equiv \frac{1}{2}\langle\chi\rmi\sigma_2\chi\rangle
  = \frac{\pi}{4}\rho_D(0) \,,
\label{eq:condensate_st}
\end{equation}
where the extra $1/2$ factor in the diquark relation comes from the
square-root prescription necessary to cancel the doubled Nambu-Gor'kov
basis.  In the above we have chosen the same normalization as
Ref.~\cite{Nishida:2003uj}.

\FIGURE{
\includegraphics[width=7cm]{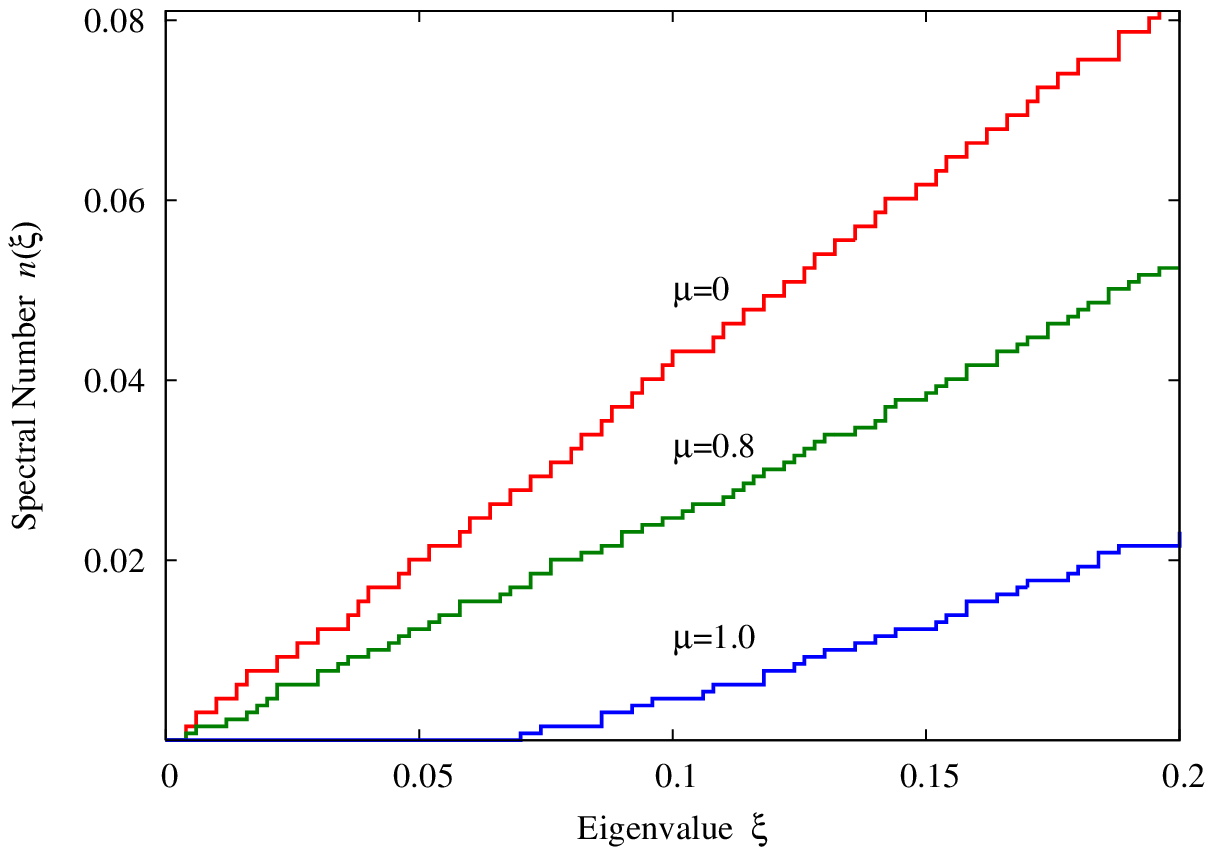}
\includegraphics[width=7cm]{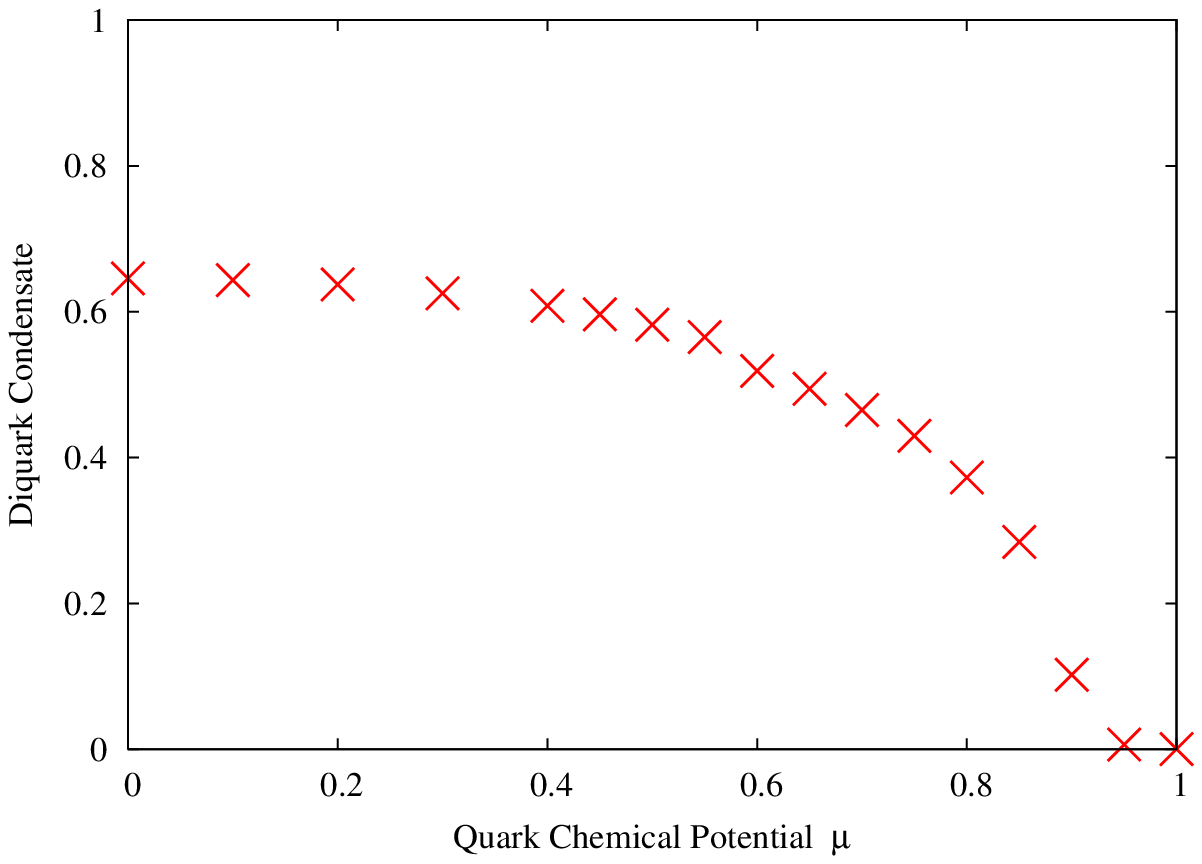}
\caption{Left) Histogram of $n_D(\xi)$ whose slope gives the spectral
  density $\rho_D(\xi)$.  Right) Diquark condensate as a function of
  $\mu$ at $\mq=0$.}
\label{fig:hist}
}

  It is intriguing to evaluate $n_D(\xi)$ by the explicit numerical
calculation for the eigenvalues in Fig.~\ref{fig:chiral_st} from which
we can get $\rho_D(\xi)$.  Figure~\ref{fig:hist} shows our results in
the chiral limit.  In this case only the diquark condensate is a
non-vanishing quantity~\cite{Dagotto:1986gw}.  We plot the diquark
condensate in the right of Fig.~\ref{fig:hist} without indicating the
error bar.  We did so because, though the fitting error is small, the
systematic error is large.  If we change the working procedure to
measure the slope from the histogram in the left of
Fig.~\ref{fig:hist}, the resultant diquark condensate would change
too.  For clarity of our numerical procedure we explain how we compute
the slope of $n_D(\xi)$ at the origin.  We assume a functional form
$n_D(\xi)=a\xi+b\xi^2$ within the range $\xi\in[0,0.1]$ and fix $a$
and $b$ to fit the data.  Then, $a$ gives the slope at the origin.
If $a$ turns negative, that means no spectral density at the origin,
and so the diquark condensate should be zero.  In this way we draw the
right of Fig.~\ref{fig:hist} which shows outstanding agreement with
the upper-left of Fig.~1 in Ref.~\cite{Nishida:2003uj}.

  The $\mq$ dependence in $\D(\mu)\D^\dagger(\mu)$ is not such trivial
as in the case of $\D(\mu)$.  Roughly speaking, a finite $\mq$ shifts
the eigenvalue in the positive real direction so that the eigenvalue
distribution is blocked in the vicinity of the origin as long as $\mu$
is small.  For $\mu$ above a certain threshold value the diquark
spectral density becomes finite at $\xi=0$, and the diquark
condensation is activated.  We can repeat the calculation in the
massive case as well.  As we mentioned our choice is $\mq=0.2$, and we
read the chiral and diquark condensate from the chiral and diquark
spectral density.  Our final results are presented below in
Fig.~\ref{fig:chiral_diquark}.  We note that the onset for the chiral
condensate decrease is determined by the front edge of the sidling
wall which corresponds to the edge of the Dirac eigenvalue
distribution, which in turn corresponds to the diquark onset.

\FIGURE{
\includegraphics[width=7cm]{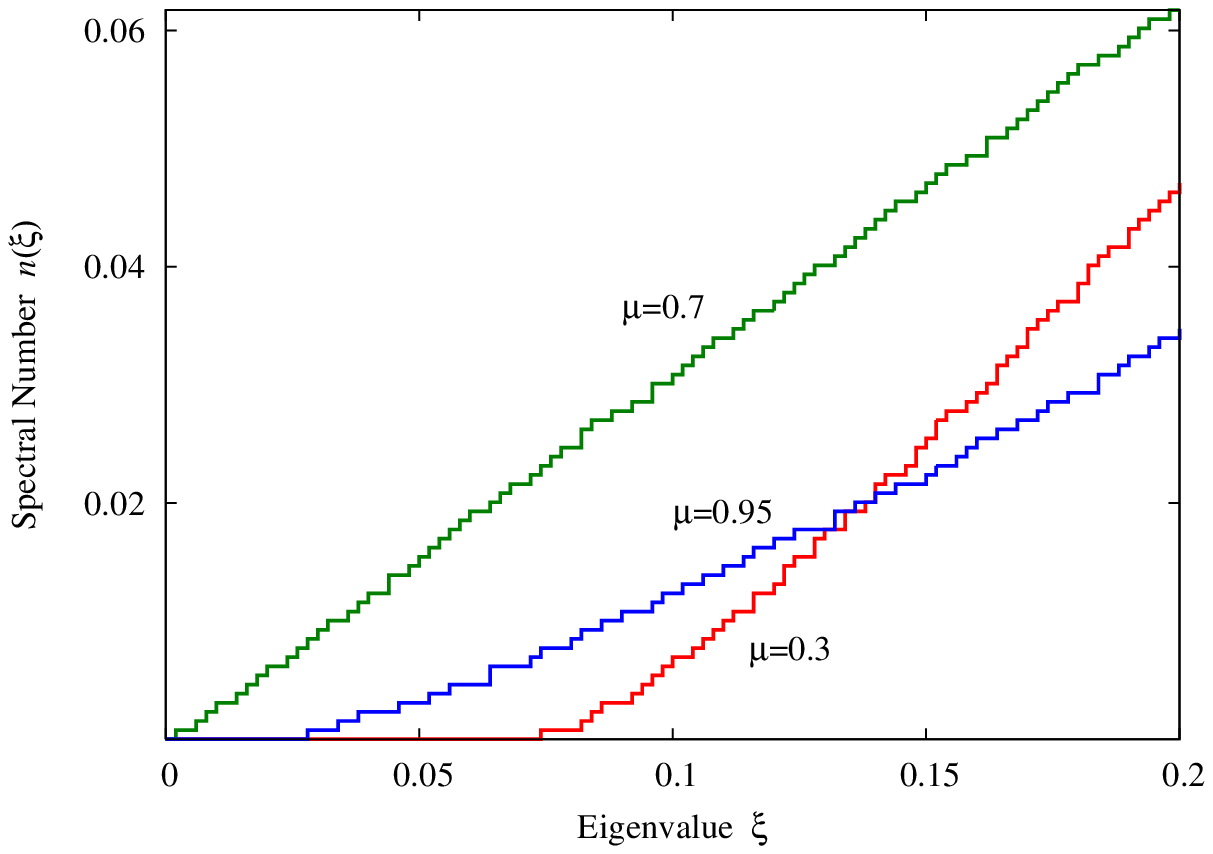}
\includegraphics[width=7cm]{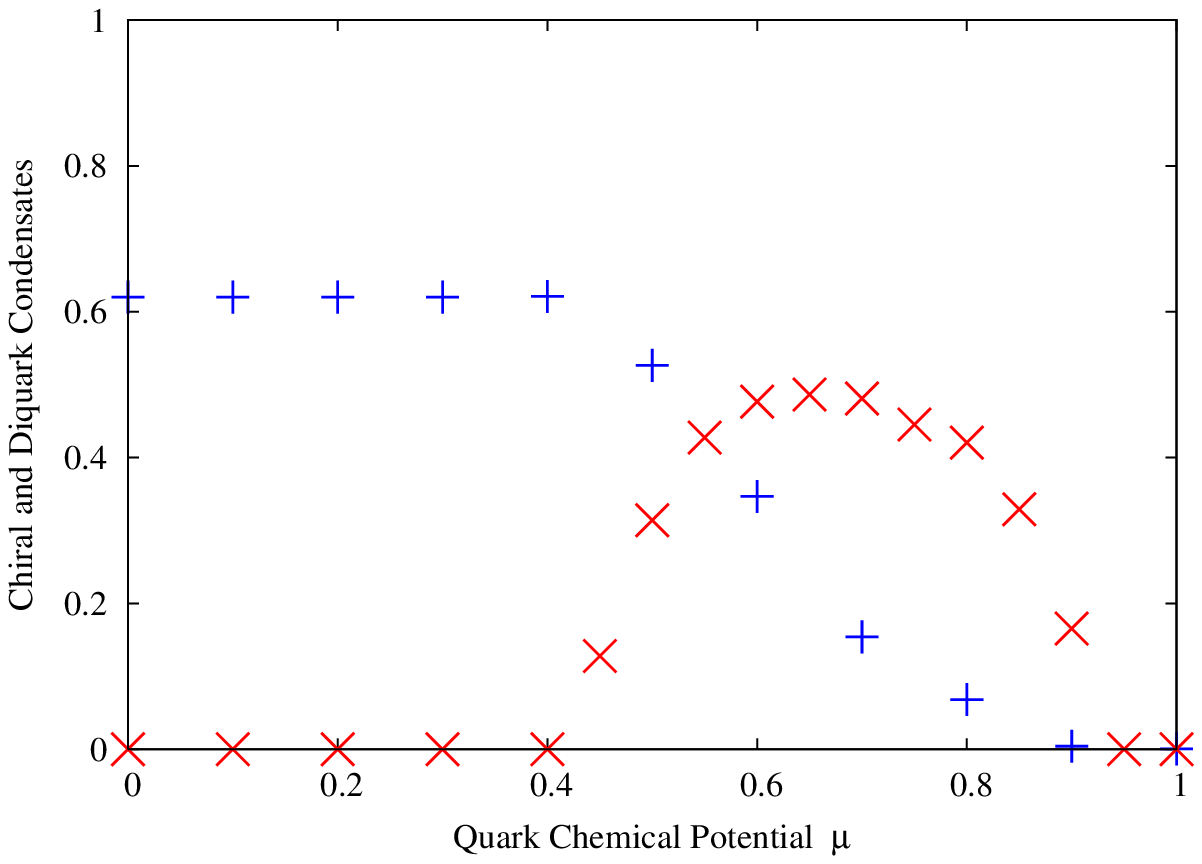}
\caption{Left) Histogram of $n_D(\xi)$ in the case of $\mq=0.2$.
  Right) Chiral and diquark condensates as a function of $\mu$ at
  $\mq=0.2$.}
\label{fig:chiral_diquark}
}

  It is impressing that the results in the right of
Fig.~\ref{fig:chiral_diquark} is consistent qualitatively with the
mean-field analysis in the strong coupling limit given in the
upper-left of Fig.~1 in Ref.~\cite{Nishida:2003uj}, though the direct
comparison is not possible for different mass choice.


\subsection{Wilson Fermion}

  We shall consider the Wilson fermion henceforth.  The Dirac operator
is defined as
\begin{equation}
 \begin{split}
 \D_W(\mu) &\equiv \delta_{m,n}-\kappa\sum_i \Bigl[(r-\gamma_i)\,U_i(m)
  \,\delta_{m+\hat{\imath},n} + (r+\gamma_i)\,U^\dagger_i(n)\,
  \delta_{m,n+\hat{\imath}} \Bigr] -\\
 &\quad -\kappa\Bigl[ (r-\gamma_4)\,\rme^\mu\,U_4(m)\,\delta_{m+\hat{\imath},n}
  + (r+\gamma_4)\,\rme^{-\mu}\,U^\dagger_4(n)\,\delta_{m,n+\hat{\imath}}
  \Bigr] \,,
 \end{split}  
\end{equation}
where $\kappa$ is the hopping parameter and we choose $r=1$ throughout
this work.  In this case we adopt $V=4^4$ and then there are
(2~colors)$\times$(4~spinors)$\times 4^4=2048$ eigenvalues.  Of
course, we could treat $V=6^4$ without difficulty, but there are then
many eigenvalues (almost five times more than the $V=4^4$ case) and
plotting looks too dense.  Our small lattice volume is limited not for
technical reason but for presentation convenience.

  It is instructive to see the free dispersion relations first.  With
the free background (i.e.\ $U_\mu=\1$ everywhere) it is easy to
calculate the eigenvalue analytically in momentum space to find
\begin{equation}
 \begin{split}
  \Re(\lambda_{\text{free}}) &= 1-2\kappa r\bigl( \cos p_1
  +\cos p_2 + \cos p_3 + \cos p_4 \bigr) \,,\\
  \Im(\lambda_{\text{free}}) &= \pm 2\kappa\sqrt{
  (\sin p_1)^2 + (\sin p_2)^2 + (\sin p_3)^2 + (\sin p_4)^2} \,.
 \end{split}
\label{eq:free}
\end{equation}
Although this expression is valid only for the free background, it
turns out to be quite useful to understand the eigenvalue distribution
in a qualitative level even at strong coupling as we will see
shortly.

  Usually $\Re(\lambda)<0$ gives the condition that the Aoki phase
appears.  In the free case, therefore, the Aoki phase has a window
$|\kappa|>1/(8r)$, while the Aoki phase condition is $|\kappa|>1/(4r)$
in the strong coupling limit.  Now, as we mentioned in Introduction, it
is confusing that the diquark condensation has exactly the same
criterion for the onset, as demonstrated in Figs.~\ref{fig:staggered}
and \ref{fig:chiral_st}.  Then, a question arises; which of the
diquark superfluid phase and the Aoki phase is more favored?  The rest
of this paper will be devoted to answering this question.

\FIGURE{
\includegraphics[width=12cm]{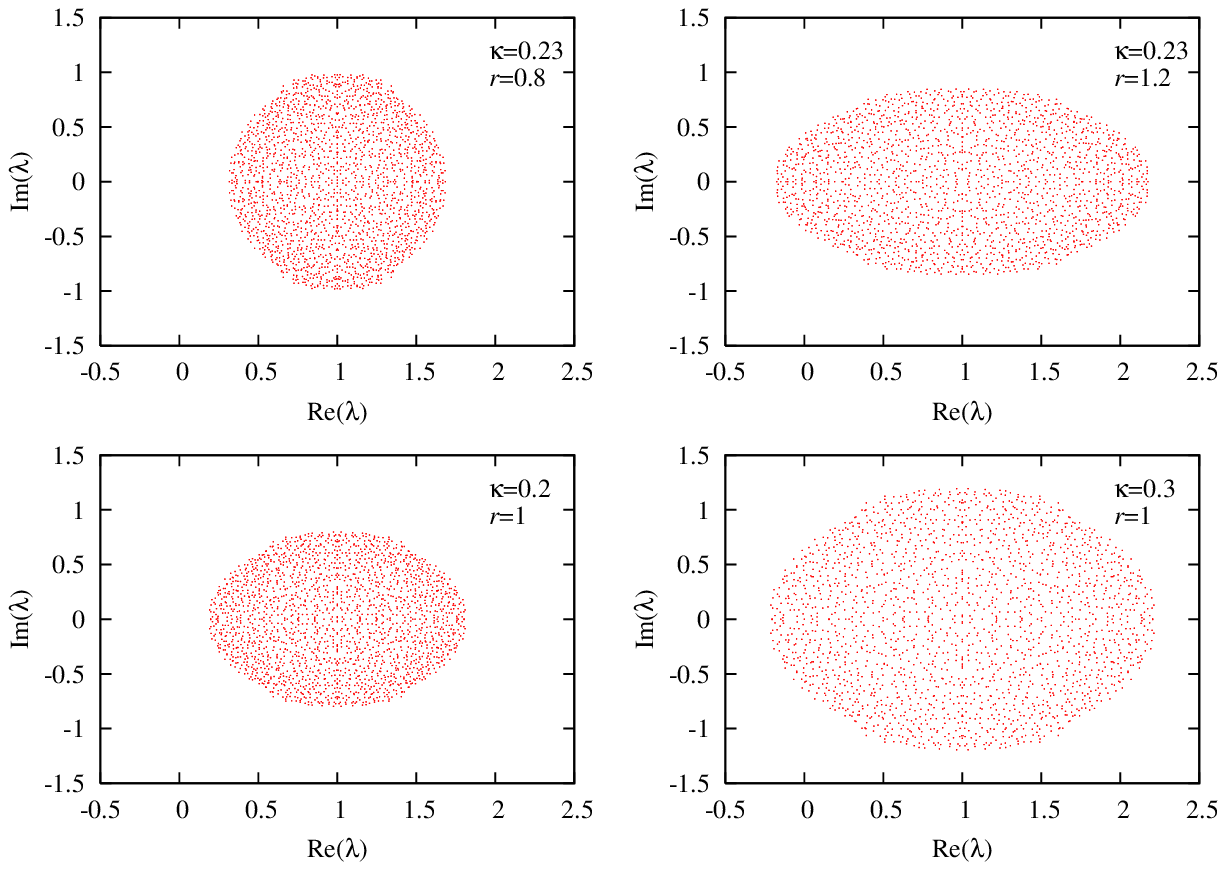}
\caption{Eigenvalue distribution for a random gauge configuration in
  the Wilson fermion formalism at $\mu=0$ on the $4^4$ lattice for
  various combinations of $\kappa$ and $r$.}
\label{fig:mu00}
}

  Let us see the parameter dependence of the eigenvalue distribution
for a random configuration in the zero density case, which is shown in
Fig.~\ref{fig:mu00}.  When we increase $\kappa$ with $r$ fixed as
shown in the lower two figures in Fig.~\ref{fig:mu00}, the
distribution range is enlarged.  We can understand this qualitatively
from Eq.~(\ref{eq:free}) in the free dispersion; both $\Re(\lambda)-1$
and $\Im(\lambda)$ are proportional to $\kappa$.  The upper two plots
in Fig.~\ref{fig:mu00} show the $r$ dependence with $\kappa$ fixed.
In this case in turn the distribution stretches only along the real
axis.  This feature is also manifest in Eq.~(\ref{eq:free}) since only
$\Re(\lambda)-1$ is multiplied by $r$.  As we can see, the
distribution penetrates into the negative real region between $(r=1,
\kappa=0.2)$ and $(r=1, \kappa=0.3)$, which is consistent with the
known fact that the critical coupling is $(r=1, \kappa=0.25)$ in the
strong coupling limit.  In what follows we will employ a value of
$\kappa=0.23$ which is close to the critical point but still outside
of the Aoki phase region, if any.

  Next, we will investigate how the chemical potential affects the
eigenvalue distribution.  Let us consider the free case first again
in which the fourth component is replaced as $p_4\to p_4-\rmi\mu$.
Then, the real and imaginary parts in the free dispersion are,
respectively, modified by
\begin{equation}
 \begin{split}
  \cos(p_4-\rmi\mu) &= \cosh\mu \cos p_4 + \rmi\sinh\mu \sin p_4 \,,\\
  \bigl[ \sin(p_4-\rmi\mu) \bigr]^2 &= (\sin p_4)^2 - (\sinh\mu)^2
   -{\textstyle\frac{1}{2}}\rmi \sinh(2\mu) \sin(2p_4) \,.
 \end{split}
\end{equation}
In this simple case it is interesting to see how the free known
results are affected by the effect of the finite chemical potential
which we show in Fig.~\ref{fig:free}.

\FIGURE{
\includegraphics[width=12cm]{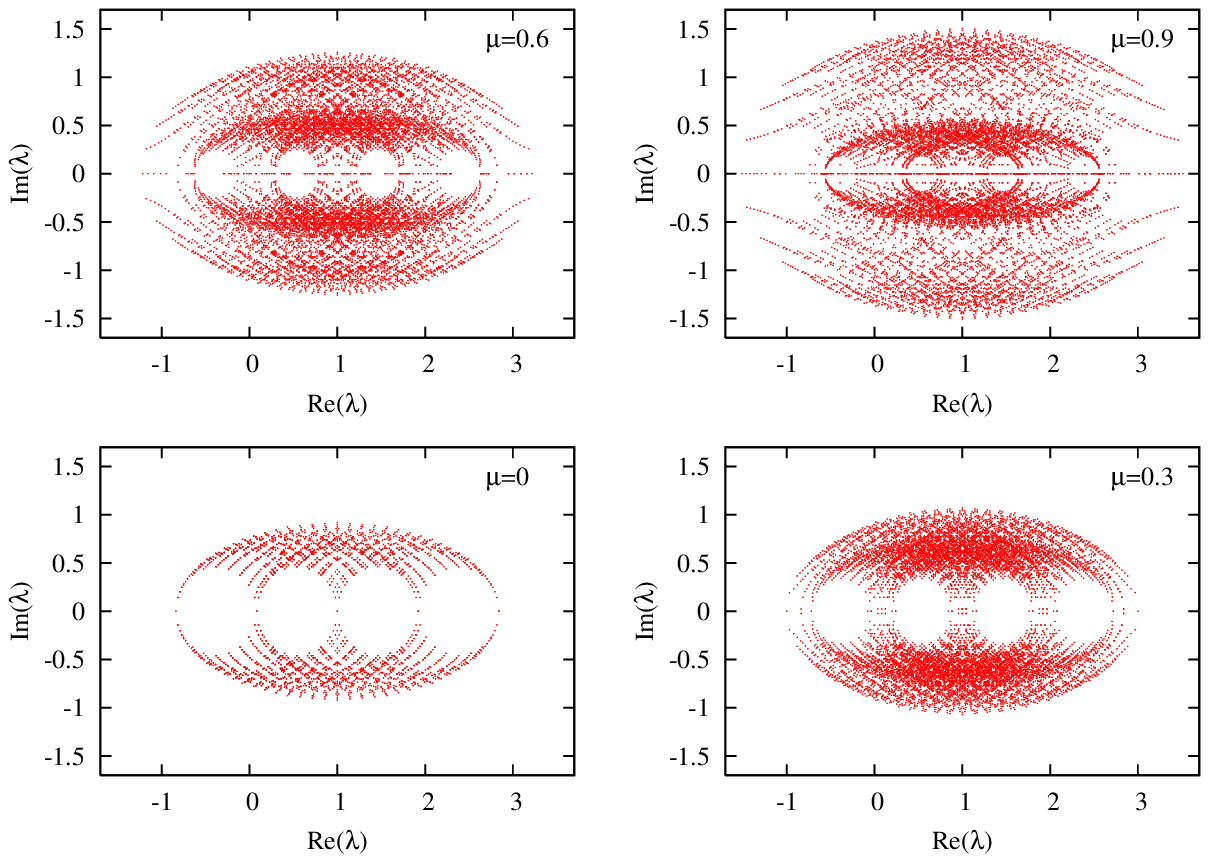}
\caption{Eigenvalue distribution for the free Wilson fermion at
  $\kappa=0.23$ on the $4^4$ lattice for $\mu=0$, $0.3$, $0.6$, and
  $0.9$.}
\label{fig:free}
}

  To draw Fig.~\ref{fig:free} we have discretized the momenta $p_1$,
$p_2$, $p_3$, and $p_4$ in the range $[-\pi,\pi]$ into twenty points
with equal spacing.  We did so in order to make the ``density''
perceivable from Fig.~\ref{fig:free};  if the momentum is close to
continuum with many points, the distribution except for the $\mu=0$
case does not have the empty region strictly.  The concentration would
be hard to see.  It is quite interesting to observe a non-trivial
structure emerging at high $\mu$ unexpectedly.  It is apparent that
the density has a similar effect as the hopping parameter $\kappa$;
the eigenvalue profile becomes wider in the complex plane as either
$\mu$ or $\kappa$ gets greater.  The density modification is not such
simple, however, and we presume that the rich contents in dense quark
matter are attributed in part to this structural difference.

  At the same time, however, we have to keep in mind that this
complicated structure at large $\mu$ looks like coming from the
non-trivial entanglement between different doubler sectors.  In the
vicinity of the continuum limit at $\mu=0$ only the far left edge part
corresponds to the lightest physical excitation and four other points
crossing the real axis are doublers going to infinity.  This clear
separation is missing in view of the eigenvalue distribution of
Fig.~\ref{fig:free} at $\mu=0.6$ or at $\mu=0.9$ for instance.  This
poses a serious question;  even though we could solve the sign problem
somehow, it should be a subtle issue how to separate the doublers out
at density high enough to allow for excitations of unphysical
doublers.

\FIGURE{
\includegraphics[width=12cm]{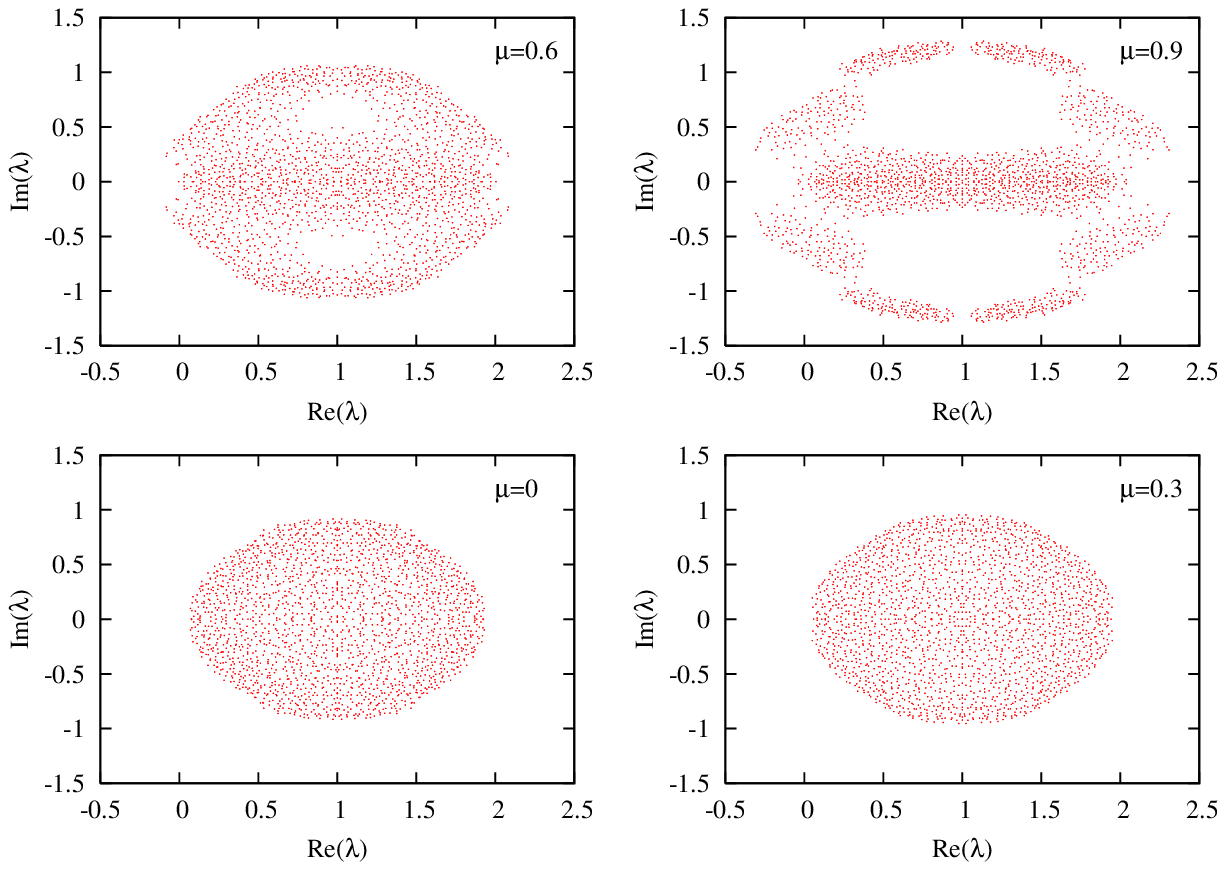}
\caption{Eigenvalue distribution for a random gauge configuration in
  the Wilson fermion formalism at $\kappa=0.23$ on the $4^4$ lattice
  for $\mu=0$, $0.3$, $0.6$, and $0.9$.}
\label{fig:wilson}
}

  For a randomly generated gauge configuration the $\mu$ dependence of
the eigenvalue distribution reflects the above mentioned structure as
displayed in Fig.~\ref{fig:wilson}.  Needless to say, we can follow
the same path to evaluate the chiral condensate but the resulting
condensate is finite and almost constant independent of the density.
This is because naive chiral symmetry is explicitly broken in the
Wilson fermion formalism, and so we will not present the results.  Let
us now evaluate the diquark condensate in the same way as we did in
the staggered fermion formalism.  It is calculable from the integrated
spectral number $n_D(\xi)$ for the operator $\D(\mu)\D^\dagger(\mu)$.
Our results are shown in Fig.~\ref{fig:diquark_w}.

\FIGURE{
\includegraphics[width=7cm]{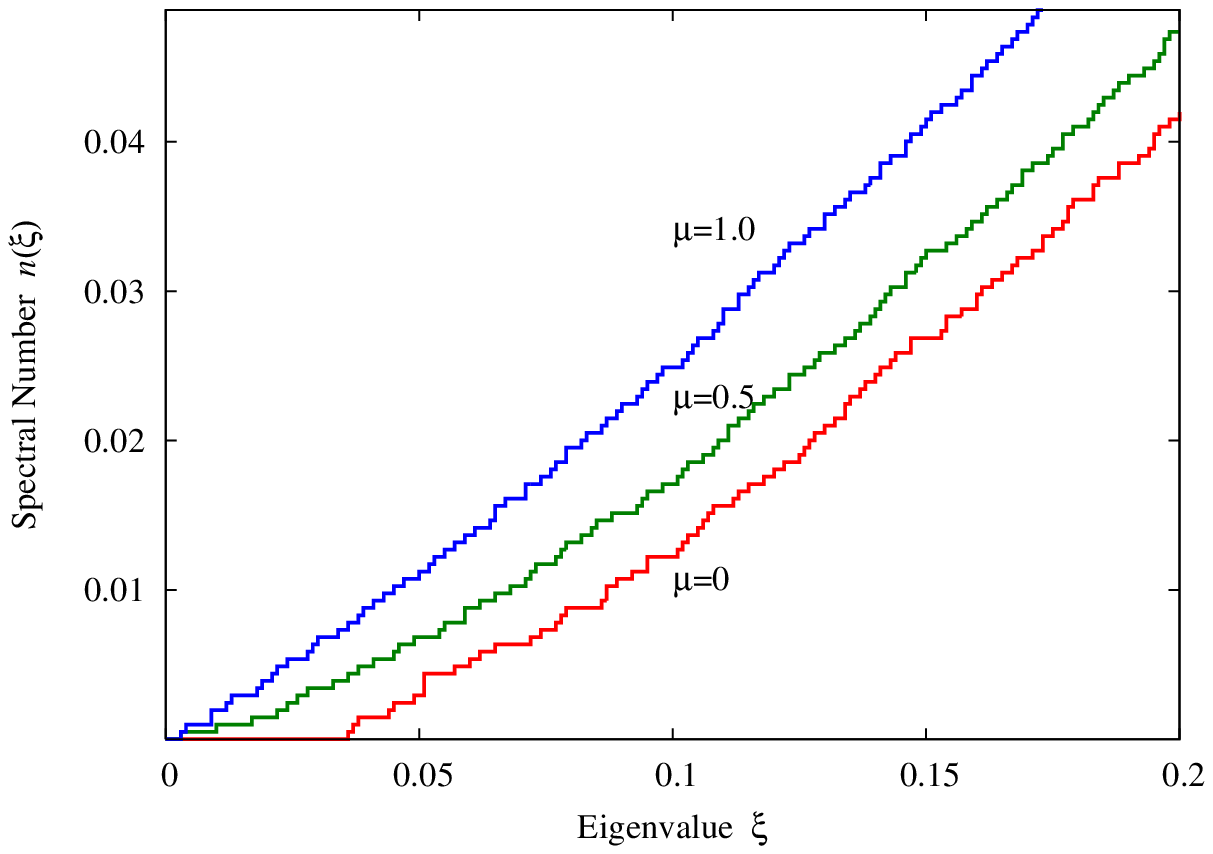}
\includegraphics[width=7cm]{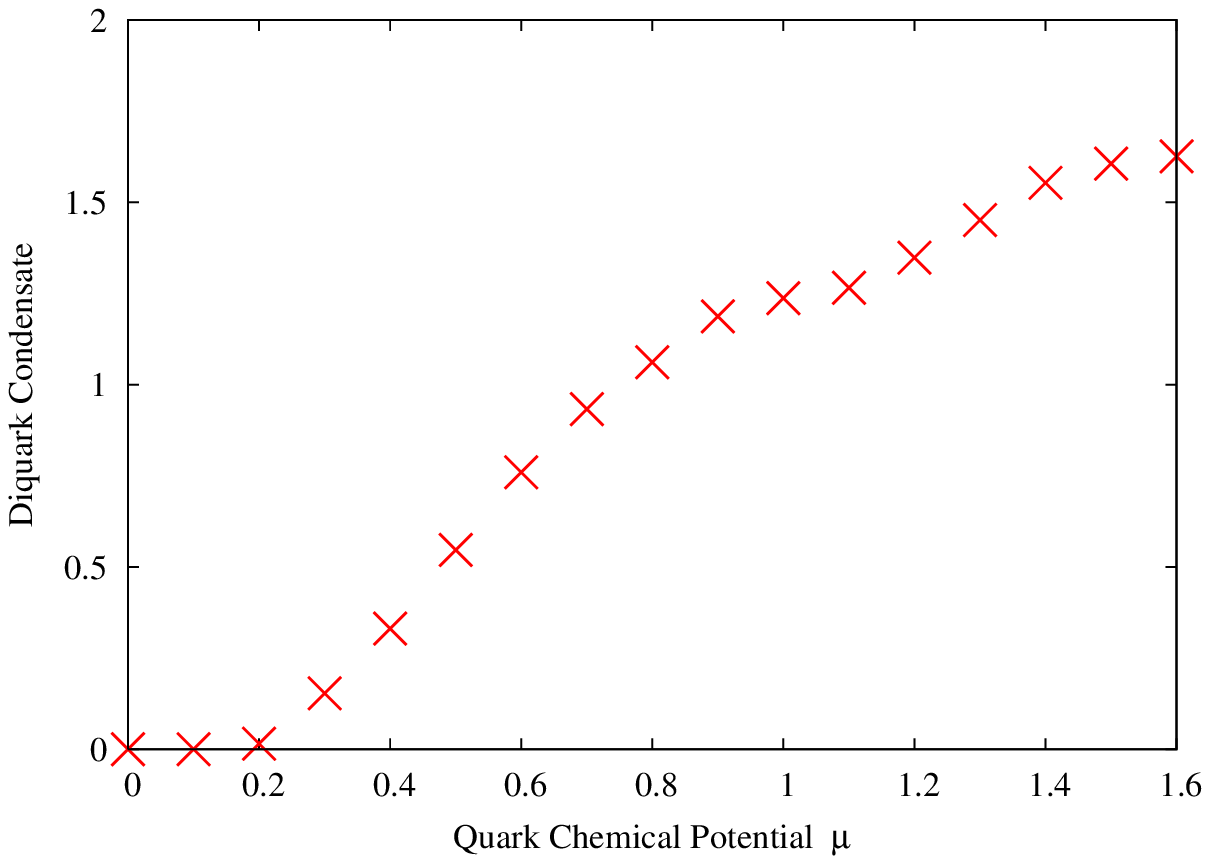}
\caption{Left) Histogram of $n_D(\xi)$.  Right) Diquark condensate as a
  function of $\mu$.}
\label{fig:diquark_w}
}

  In the case of the Wilson fermion there are four times degrees of
freedom than the staggered fermion and so the saturation effect is not
yet relevant in the right of Fig.~\ref{fig:diquark_w}.  It should be
mentioned that we measure the slope at the origin in the same way as
in the staggered fermion;  we fit the data up to $\xi=0.1$ by
$n_D(\xi)=a\xi+b\xi^2$.  If we change the fitting range and the
fitting functional form, we would have quantitatively different
results.  The systematic error is not well under control.  At least,
however, we can state that such simple calculations in this work could
capture qualitative features of diquark superfluidity.

  Finally let us discuss the possibility of the Aoki phase from the
point of view of the spectral density $\rho_H(\eta)$ corresponding to
the parity-flavor breaking condensate.  That is understood from the 3D
plot for the spectral density given in Fig.~\ref{fig:aoki}.

\FIGURE{
\includegraphics[width=12cm]{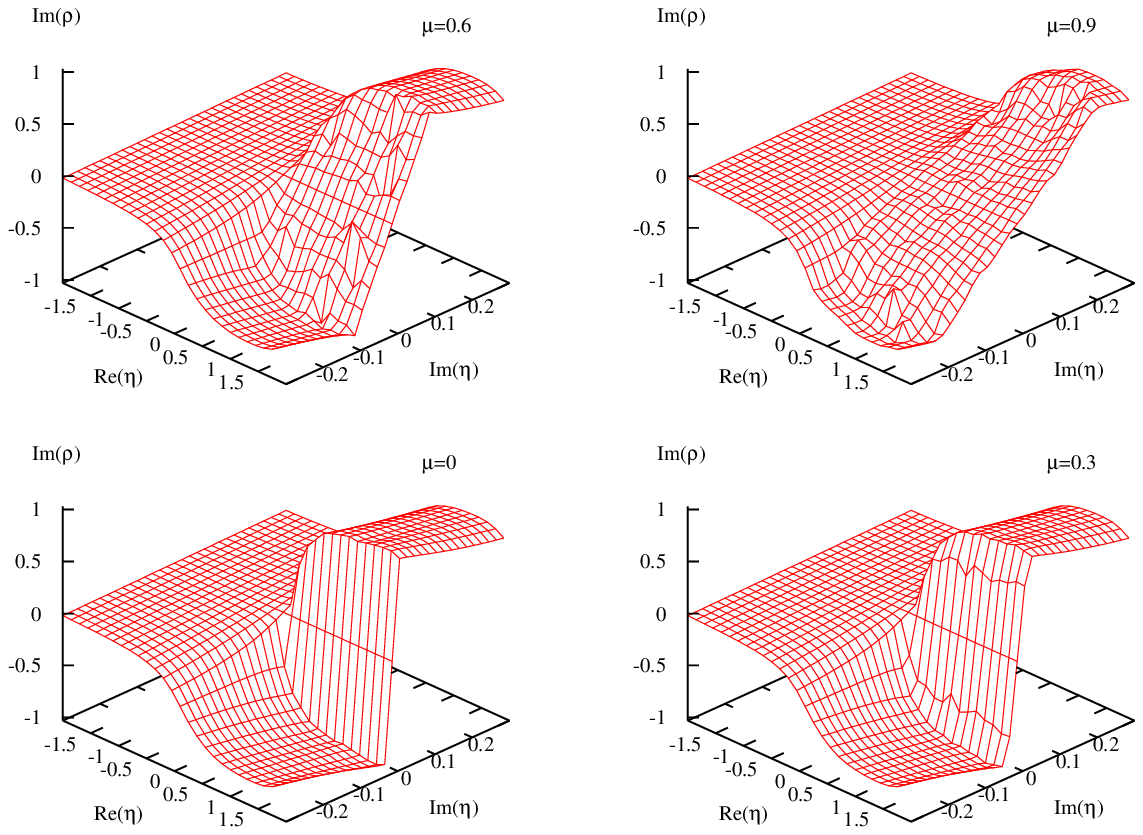}
\caption{Imaginary part of the spectral density, $\Im(\rho_H(\eta))$
  in the complex plane for various values of the chemical potential.}
\label{fig:aoki}
}

  According to the Banks-Casher-type relation we obtained, the
parity-flavor breaking condensate is to be acquired from the height of
$\Im(\rho_H(\eta))$ at the origin.  It should be the finite-volume
effect that the origin looks smooth and the symmetry breaking looks
like not occurring even at $\mu=0$.  We anticipate that the standing
wall would be more sharp upright around the origin in the
thermodynamic limit.  Even in the thermodynamic limit, however, the
wall has a finite slope at $\mu\neq0$ which reminds us of the chiral
condensate discussed in Fig.~\ref{fig:chiral_st}.  Thus, the same
conclusion can be drawn;  the parity-flavor breaking condensate thus
takes a non-zero value in the limit of $H\to0$ while keeping $\mu=0$
strictly.  In the presence of infinitesimal chemical potential, in
contrast, the situation changes and the condensate is vanishing in the
limit of $\mu\to0$ after taking the limit $H=0$.  Therefore, in the
exactly same sense as the chiral dynamics we should conclude that
there is no parity-flavor breaking condensate in this system.  As long
as the two-color QCD with two degenerate flavors is concerned, we do
not have to care about the Aoki phase even in the Wilson fermion
formalism on the lattice.


\section{Remarks}

  We saw the eigenvalue distribution of the Dirac operator at finite
density $\D(\mu)$ and its relatives $\D(\mu)\D^\dagger(\mu)$ and
$\D(\mu)D^\dagger(-\mu)$ to discuss the fate of the chiral condensate,
the diquark condensate, and the parity-flavor breaking condensate.

  We have a conjecture that a similar pattern in the eigenvalue
distribution should appear also in dense QCD with three colors;  the
eigenvalue distribution reaches the origin at the onset for nuclear
matter.  At this stage we have no idea what kind of characteristic
feature is associated with the color superconducting phase.  We
believe, in principle, that we can pursue our strategy in order to
access color superconductivity.  It has not been successful so far to
describe the color superconducting phase in the strong coupling
limit~\cite{Azcoiti:2003eb}.  Since our method does not assume any
mean-field nor truncation, our method should be useful to clarify what
is going on in the diquark channel in strong-coupling QCD.\ \ This is
what we are planning to do as a future extension.

  The present work is focused on the numerical outputs.  It is maybe
an interesting question how the change in the eigenvalue distribution
could be interpreted in analogy to known phenomena such as the chiral
symmetry breaking interpreted as the Anderson
localization~\cite{GarciaGarcia:2006gr,Takahashi:2007dv,Takahashi:2008mm}.
This research deserves further investigation.  Also, it should be
feasible, in principle, to apply our idea to the weak-coupling regime
close to the continuum limit using the open gauge configurations if
they are available.  Since the physical units in the color SU(2) world
are obscure, unfortunately, the continuum limit is not quite lucid
then.  Nevertheless, in view of the qualitative success of the strong
coupling expansion to understand hot and dense
QCD~\cite{Fukushima:2003vi}, we may well anticipate that a smooth
crossover links the strong-coupling regime to the weak-coupling one.
This could be checked by inclusion of the finite $\beta$ corrections.

  Finally let us mention on the possible extension to the overlap
fermion where exact chiral symmetry can be defined on the lattice.
Then, there is no need to consider the Aoki phase from the beginning
because the eigenvalue distribution sits on a single circle line at
$\mu=0$.  This nice feature breaks down, however, at finite density.
This is because the $\gamma_5$-hermiticity is lost at $\mu\neq0$ but
it is amazing that chiral symmetry is still
realized~\cite{Bloch:2006cd}.  The overlap fermion surpasses the
staggered and Wilson fermions;  we would be able to treat not four but
two flavors and look into the behavior of the chiral condensate as
well as the diquark condensate in the overlap fermion formalism.  This
extension is also on our list for future perspective.

\acknowledgments
  The author thanks the colleagues of Yukawa Institute for
Theoretical Physics and of Department of Physics at the Kyoto
University, especially Toru~T.~Takahashi and Hideaki~Iida, for useful
conversations.  He also thanks Atsushi~Nakamura for discussions.  This
work is in part supported by Yukawa International Program for
Quark-Hadron Sciences.


\end{document}